\documentclass[prl,twocolumn,twoside,showpacs,nofootinbib,floatfix]{revtex4}

\usepackage{graphicx,color}
\usepackage{amsmath,amssymb,bm}
\usepackage{times}

\newcommand{\eq}[1]{\begin{equation}#1\end{equation}}

\newcommand{\ket}[1]{\ensuremath{\,|{#1}\rangle}}

\newcommand{\matrixe}[3]{\ensuremath{\langle{#1}|\,{#2}\,|{#3}\rangle}}

\newcommand{\op}[1]{\ensuremath{\mathrm{#1}}}

\newcommand{\HO}{\ensuremath{\op{H}}}
\newcommand{\TO}{\ensuremath{\op{T}}}
\newcommand{\VO}{\ensuremath{\op{V}}}

\newcommand{\UCOM}{\ensuremath{\textrm{UCOM}}}
\newcommand{\elem}[2]{\ensuremath{{}^{#2}\text{#1}}}

\newcommand{\symboldiamond}[1][black]{{\color{#1}$\blacklozenge$}}

\newcommand{\symbolbox}[1][black]{{\color{#1}$\blacksquare$}}
\newcommand{\symbolcircle}[1][black]{{\color{#1}$\bullet$}}

\definecolor{FGViolet}{rgb}{0.61,0.32,0.61}
\definecolor{FGBlue}{rgb}{0,0,0.8}
\definecolor{FGGreen}{rgb}{0.2,0.7,0.2}
\definecolor{FGRed}{rgb}{0.8,0,0}
\definecolor{FGWhite}{rgb}{1,1,1}
\definecolor{FGGray}{rgb}{0.5,0.5,0.5}
\definecolor{FGBlack}{rgb}{0,0,0}

\begin{document}

\title{Ab Initio Study of \elem{Ca}{40} with an Importance Truncated No-Core Shell Model}

\author{R. Roth}
\email{robert.roth@physik.tu-darmstadt.de}
\affiliation{Institut f\"ur Kernphysik, TU Darmstadt, Schlossgartenstr. 9,
64289 Darmstadt, Germany}

\author{P. Navr\'atil}
\email{navratil1@llnl.gov}
\affiliation{Lawrence Livermore National Laboratory, P.O. Box 808, L-414, Livermore, CA 94551, USA}

\date{\today}

\begin{abstract}  
We propose an importance truncation scheme for the no-core shell model, which enables converged calculations for nuclei well beyond the p-shell. It is based on an \emph{a priori} measure for the importance of individual basis states constructed by means of many-body perturbation theory. Only the physically relevant states of the no-core model space are considered, which leads to a dramatic reduction of the basis dimension. We analyze the validity and efficiency of this truncation scheme using different realistic nucleon-nucleon interactions and compare to conventional no-core shell model calculations for \elem{He}{4} and \elem{O}{16}. Then, we present the first converged calculations for the ground state of \elem{Ca}{40} within no-core model spaces including up to $16\hbar\Omega$-excitations using realistic low-momentum interactions. The scheme is universal and can be easily applied to other quantum many-body problems.
\end{abstract}


\maketitle

%%%%%%%%%%%%%%%%%%%%%%%%%%%%%%%%%%%%%%%%%%%%%%%%%%%%%%%%%%%%%%%%%%%%%%
%%%%%%%%%%%%%%%%%%%%%%%%%%%%%%%%%%%%%%%%%%%%%%%%%%%%%%%%%%%%%%%%%%%%%%
%%%%%%%%%%%%%%%%%%%%%%%%%%%%%%%%%%%%%%%%%%%%%%%%%%%%%%%%%%%%%%%%%%%%%%
\clearpage

The theoretical description of nuclei as strongly interacting quantum many-body systems bears two fundamental challenges: (i) The interaction between the constituents is a complicated, residual force resulting from the strong interaction of quantum chromo dynamics (QCD). (ii) The nuclear many-body problem, which is dominated by correlations, has to be solved reliably for systems containing from 2 to a few hundred nucleons. There has been substantial progress on both aspects over recent years. In addition to traditional realistic nucleon-nucleon interactions, like the Argonne V18 \cite{WiSt95} and the CD Bonn potential \cite{Mach01}, which reproduce the experimental nucleon-nucleon scattering data with high precision, interactions derived in the framework of a chiral effective field theory became available \cite{EnMa03,EpNo02}. The latter provide a systematic and consistent picture of two- and three-nucleon interactions based on the symmetries of QCD. Given these interactions the solution of the nuclear many-body problem remains a formidable task. For light nuclei, \emph{ab initio} methods like the no-core shell model \cite{NaVa00,NaOr02} or the Green's Function Monte Carlo method \cite{PiWi01,PiWi04} provide tools to solve the many-body problem numerically. Recently, the former has been used to study the properties of mid-p-shell nuclei based on chiral two- plus three-nucleon interactions for the first time \cite{NaGu07}. 

Methods like the full no-core shell model (NCSM) become computationally intractable for heavier nuclei. Currently converged NCSM calculations are limited to the p-shell. For heavier nuclei one relies on severe approximations, such as perturbative expansions on top of a Hartree-Fock solution \cite{RoPa06,BaPa06}, or completely phenomenological approaches. Only few approaches, e.g. the coupled-cluster method \cite{KoDe04,WlDe05}, are able connect the domain of strict \emph{ab initio} methods with the domain of approximate models for heavier systems. 

In order to extend the scope of the \emph{ab initio} NCSM to heavier nuclei we propose an extension using a priori information on the importance of individual basis states obtained from many-body perturbation theory. We use this importance measure to truncate the NCSM model space to the physically relevant basis states and thus reduce its dimension substantially. The current limitation of full NCSM calculations results from the sheer dimension of the matrix eigenvalue problem. Typically the many-body basis, which is represented by Slater determinants of harmonic oscillator single-particle states, is truncated at a maximum number $N_{\max}$ of harmonic-oscillator excitation quanta thus defining the so-called $N_{\max}\hbar\Omega$ model space. Its (m-scheme) dimension grows exponentially with the number of nucleons $A$ and the truncation level $N_{\max}$. For \elem{O}{16} this limits the presently tractable model space to $N_{\max}=8$ \cite{CaMa05} corresponding to an effective m-scheme dimension of $6\times10^{8}$. For \elem{Ca}{40} the dimension of the 8$\hbar\Omega$ model space is $2 \times10^{12}$---well beyond the capabilities of current shell model codes. Many of these basis states are irrelevant for the description of any particular eigenstate, e.g. the ground state. Therefore, if one were able to identify the important basis states beforehand, one could reduce the dimension of the matrix eigenvalue problem without loosing predictive power. This, as we propose in this Letter, can be done using many-body perturbation theory. This type of problem is not restricted to nuclear physics but common to many areas of quantum many-body theory. For the description of strongly corelated lattice systems, e.g. in the context of ultracold atomic gases in optical lattices \cite{ScHi07}, or other atomic or condensed matter systems, one encounters the same limitations of exact diagonalization approaches. 

We start out with a translationally invariant Hamiltonian $\HO_{\text{int}}$ composed of the intrinsic kinetic energy $\TO_{\text{int}}=\TO-\TO_{\text{cm}}$ and a two-nucleon interaction $\VO_{\text{NN}}$. In the first part of this  Letter, we use the phase-shift equivalent correlated interaction $\VO_{\UCOM}$ discussed in Refs. \cite{RoHe05,RoPa06}. For three- and four-nucleon systems it provides binding energies in good agreement with experiment without the inclusion of an explicit three-nucleon force. Together with its good convergence properties it provides a suitable benchmark interaction for the present study. In the second part we use the $V_{\text{low}k}$ low-momentum interaction \cite{BoKu03} to discuss the ground state structure of \elem{O}{16} and \elem{Ca}{40}.

The general concept of the importance truncation is as follows: We start with a reference state $\ket{\Psi_{\text{ref}}}$, which represents a zeroth-order approximation of the many-body state we are interested in. Here, we limit ourselves to the ground state, but the scheme can easily be extended to excited states or more than one reference state. Starting from  $\ket{\Psi_{\text{ref}}}$ we can build a many-body space by generating all possible $n$-particle--$n$-hole ($n$p$n$h) excitations up to the excitation energy $N_{\max}\hbar\Omega$. By increasing $n$ ($\leq A$), we eventually recover the translationally invariant $N_{\max}\hbar\Omega$ model-space of the NCSM. We now estimate the contribution of a given $n$p$n$h-state $\ket{\Phi_{\nu}}$ to the exact eigenstate after the NCSM diagonalization via many-body perturbation theory. In first-order MBPT the amplitude of the state $\ket{\Phi_{\nu}}$ is given by 
\eq{ \label{eq:importanceweight}
  \kappa_{\nu} = -\frac{\matrixe{\Phi_{\nu}}{\HO'}{\Psi_{\text{ref}}}}{\epsilon_{\nu}-\epsilon_{\text{ref}}} \;,
}
where $\HO'$ is the Hamiltonian of the perturbation and $\epsilon_{\nu}$ are the unperturbed energies of the two configurations. When using the harmonic oscillator basis, the unperturbed Hamiltonian is just the one-body harmonic oscillator Hamiltonian, $\HO_0 = \HO_{\text{HO}}$, and the perturbation is $\HO'=\HO_{\text{int}} - \HO_{\text{HO}}$, where the contribution of $\HO_{\text{HO}}$ to the matrix element in \eqref{eq:importanceweight} vanishes.
  
If we restrict ourselves to two-body interactions, then $\HO'$ contains only one- and two-body terms such that  $\kappa_{\nu}$ vanishes for 3p3h and higher-order configurations. In principle higher orders of perturbation theory are required to generate states beyond the 2p2h level directly. Since this becomes computationally inefficient, we resort to an iterative scheme. In a first iteration we generate 1p1h and 2p2h states starting from the reference state, retain those with an importance weight $\kappa_{\nu} \geq \kappa_{\min}$, and solve the eigenvalue problem in this space. In the next iteration, the dominant components of the ground state $\ket{\Psi_{0}}=\sum_{\nu} C_{\nu} \ket{\Phi_{\nu}}$ obtained from the diagonalization in the previous step are used as reference state, i.e. $\ket{\Psi_{\text{ref}}}=\sum_{\nu}^{|C_{\nu}|\geq C_{\text{ref}}} C_{\nu} \ket{\Phi_{\nu}}$ with $C_{\text{ref}}\approx0.005$. Since this state already contains up to 2p2h admixtures one obtains nonvanishing importance weights \eqref{eq:importanceweight} for states up to the 4p4h level. After applying the  importance truncation $\kappa_{\nu} \geq \kappa_{\min}$ the eigenvalue problem is solved in the extended space. This cycle can be repeated until the full $N_{\max}\hbar\Omega$ model-space is generated in the limit $\kappa_{\min}=0$ . For the first applications presented here, we restrict ourselves to the 4p4h level, extensions will be discussed elsewhere.

%%%%%%%%%%%%%%%%%%%%%%%%%%%%%%%%%%%%%%%%%%%%%%%%%%%%%%%%%%%%%%%%%%%%%%
\begin{figure}[t]
\includegraphics[width=\columnwidth]{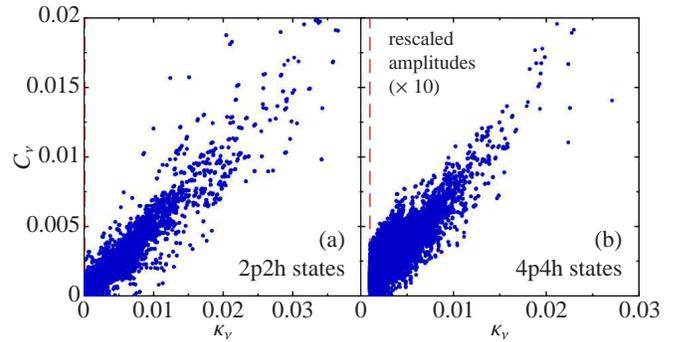}
\caption{(color online) Correlation between the \emph{a priori} importance weight $\kappa_{\nu}$ and the \emph{a posteriori} amplitudes $C_{\nu}$ resulting from the explicit diagonalization. Shown are the 2p2h (a) and 4p4h contributions (b) for a $6\hbar\Omega$ calculation for \elem{O}{16} with $V_{\text{UCOM}}$ at $\hbar\Omega=20\text{MeV}$. The dashed vertical line marks a typical truncation parameter $\kappa_{\min}$.}
\label{fig:kapcoeff_correlation}
\end{figure}
%%%%%%%%%%%%%%%%%%%%%%%%%%%%%%%%%%%%%%%%%%%%%%%%%%%%%%%%%%%%%%%%%%%%%%

The correlation plots presented in Fig. \ref{fig:kapcoeff_correlation} demonstrate that this prescription provides a reliable \emph{a priori} estimate for the size of the \emph{a posteriori} amplitudes $C_{\nu}$ of the individual basis states after solving the eigenvalue problem. There is a clear correlation between the coefficients $C_{\nu}$ and the importance measure $\kappa_{\nu}$. Therefore, the additional truncation of the many-body space to configurations with $\kappa_{\nu} \geq \kappa_{\min}$ provides an efficient means to remove irrelevant configurations from the outset. Eventually the eigenvalue problem within the truncated space is solved without further approximation. Hence, formal properties such as the variational principle and the Hylleraas-Undheim theorem remain intact---in contrast to other approaches like the coupled cluster method \cite{KoDe04,WlDe05}. When reducing the truncation parameter $\kappa_{\min}$, we are guaranteed to converge from above to the exact eigenvalues for the given interaction. 

%%%%%%%%%%%%%%%%%%%%%%%%%%%%%%%%%%%%%%%%%%%%%%%%%%%%%%%%%%%%%%%%%%%%%%
\begin{figure}[t]
\includegraphics[width=1\columnwidth]{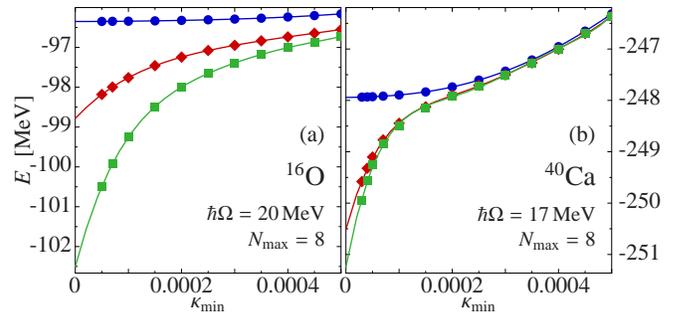}
\caption{(color online) Dependence of the ground-state energy of \elem{O}{16} (a) and \elem{Ca}{40} (b) on the truncation parameter $\kappa_{\min}$ obtained with $\VO_{\UCOM}$. The three data sets correspond to model spaces with up to 2p2h (\symbolcircle[FGBlue]), 3p3h (\symboldiamond[FGRed]), and 4p4h states (\symbolbox[FGGreen]), respectively. The lines show a 5th order polynomial interpolation.}
\label{fig:kapdependence}
\end{figure}
%%%%%%%%%%%%%%%%%%%%%%%%%%%%%%%%%%%%%%%%%%%%%%%%%%%%%%%%%%%%%%%%%%%%%%

The dependence of the ground state energy on the truncation parameter $\kappa_{\min}$ is illustrated in Fig. \ref{fig:kapdependence} for \elem{O}{16} and \elem{Ca}{40}. The different data sets correspond to calculations with spaces comprising up to 2p2h, 3p3h, and 4p4h states, respectively. We always observe a very regular behavior, which allows for an extrapolation of the energies towards $\kappa_{\min}=0$ corresponding to the complete model space. We use a 5th order polynomial fit that turns out to be very robust.

%%%%%%%%%%%%%%%%%%%%%%%%%%%%%%%%%%%%%%%%%%%%%%%%%%%%%%%%%%%%%%%%%%%%%%
\begin{figure}
\includegraphics[width=0.8\columnwidth]{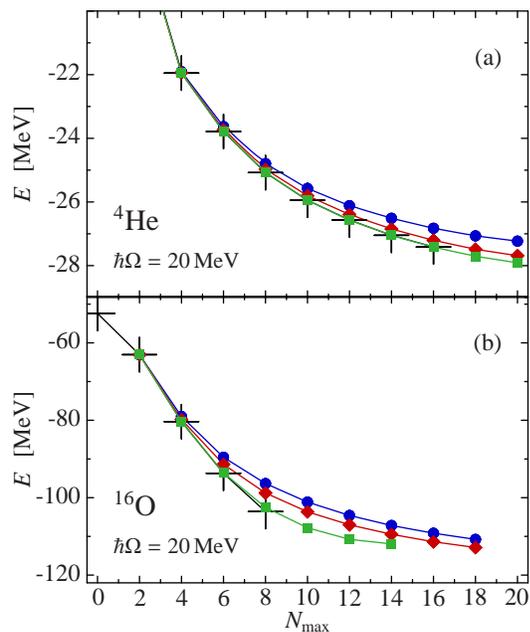}
\caption{(color online) Convergence of the ground-state energy for \elem{He}{4} (a) and \elem{O}{16} (b) versus  $N_{\max}$ obtained using $\VO_{\UCOM}$. Shown are three data sets corresponding to model spaces with up to 2p2h (\symbolcircle[FGBlue]), 3p3h  (\symboldiamond[FGRed]), and 4p4h-states  (\symbolbox[FGGreen]), respectively. Black crosses ($+$) indicate the results of full NCSM calculations. Lines to guide the eye.}
\label{fig:He4O16_Nhwdependence}
\end{figure}
%%%%%%%%%%%%%%%%%%%%%%%%%%%%%%%%%%%%%%%%%%%%%%%%%%%%%%%%%%%%%%%%%%%%%%

A first indication of the efficiency of the importance truncation scheme is given in Fig. \ref{fig:He4O16_Nhwdependence}. Shown are the ground state energies of \elem{He}{4} and \elem{O}{16} as function of $N_{\max}$ computed for the $\VO_{\UCOM}$ interaction using the importance truncation (after extrapolation to $\kappa_{\min}=0$) in comparison to the full NCSM calculation performed with the \textsc{Antoine} code \cite{CaNo99}. For \elem{He}{4} the truncated calculation including up to 4p4h configurations generates the full $N_{\max}\hbar\Omega$ model space in the limit $\kappa_{\min}=0$. The excellent agreement between full and truncated calculation in this case proves the reliability of the importance truncation scheme. Nonetheless, the dimension of the eigenvalue problem is drastically reduced by the importance truncation: For $N_{\max}=20$ the dimension of the full NCSM model space is beyond $2\times10^7$ whereas the dimension of the truncated basis is of the order $10^5$. This redcution becomes even more striking for heavier systems. 

For \elem{O}{16} a similar picture emerges, the importance truncated calculations up to the 4p4h level are in very good agreement with the full NCSM. For the $8\hbar\Omega$ model space the full calculation yields a slightly lower ground-state energy, which is due to states beyond the 4p4h level. The relevance of 4p4h correlations is a specific property of \elem{O}{16} ($\alpha$-clustering) and reveals through the compareably strong effect of those states on the ground-state energy.

%%%%%%%%%%%%%%%%%%%%%%%%%%%%%%%%%%%%%%%%%%%%%%%%%%%%%%%%%%%%%%%%%%%%%%
\begin{figure}
\includegraphics[width=0.9\columnwidth]{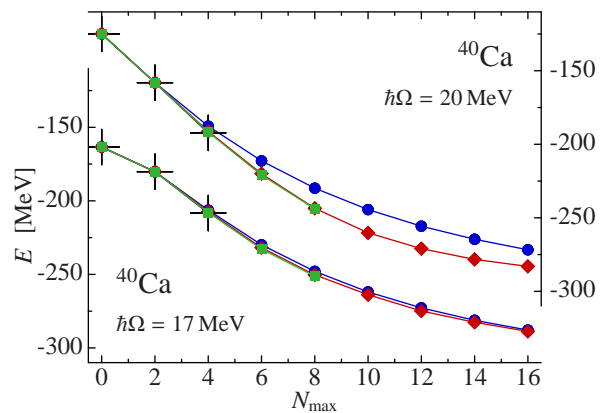}
\caption{(color online) Convergence of the ground-state energy of \elem{Ca}{40} as function of model-space size $N_{\max}$ for $\hbar\Omega=17\,\text{MeV}$ (lower curves, left-hand axis) and $20\,\text{MeV}$ (upper curves, right-hand axis) using the $\VO_{\UCOM}$ interaction. Symbols as described in Fig. \ref{fig:He4O16_Nhwdependence}.}
\label{fig:Ca40_Nhwdependence}
\end{figure}
%%%%%%%%%%%%%%%%%%%%%%%%%%%%%%%%%%%%%%%%%%%%%%%%%%%%%%%%%%%%%%%%%%%%%%

The $N_{\max}$-dependence of the ground-state energy of $\elem{Ca}{40}$ for two different oscillator frequencies is presented in Fig. \ref{fig:Ca40_Nhwdependence}. In comparison to  \elem{O}{16}, the impact of 4p4h configurations is negligible and it is expected that higher-order configurations will not affect the ground-state energy. For the oscillator frequency $\hbar\Omega=17\,\text{MeV}$ even the inclusion of 3p3h states does not lead to a significant effect. An exponential extrapolation of the 3p3h energies for $\hbar\Omega=17\,\text{MeV}$ to infinite model-space size gives $E_{\infty} \approx -316\,\text{MeV}$, which can be compared to the experimental binding energy of $-342.05\,\text{MeV}$. Keeping in mind that our calculation---because of the restriction to the 3p3h level---provides an upper bound, this result is very encouraging. It shows that the phase-shift equivalent two-nucleon interaction $\VO_{\UCOM}$ is able to provide a realistic description of binding energies also for heavier nuclei without the inclusion of an additional three-body interaction.

In this intermediate mass regime the importance truncation works very efficiently: The dimension of the importance truncated space used for \elem{Ca}{40} around $N_{\max}=16$ is of the order of $10^7$, which is smaller than the full NCSM space at $N_{\max}=4$. In order to warrant that contributions of spurious center-of-mass excitations to the energy are negligible, we monitor the expectation value of the operator $\HO_{\text{HO}}^{\text{cm}}-\frac{3}{2}\hbar\Omega$. We obtain typical values of $100\,\text{keV}$ and below proving that the importance truncation does not generate spurious center-of-mass excitations, which are otherwise absent in a complete $N_{\max}\hbar\Omega$ model space. 

Following these benchmark calculations with $\VO_{\UCOM}$, we apply the importance truncation for a systematic study of the ground states of \elem{O}{16} and \elem{Ca}{40} with the $V_{\text{low}k}$ low-momentum interaction derived from the Argonne V18 potential with a cutoff momentum $\Lambda=2.1\,\text{fm}^{-1}$ \cite{HagePriv}. The $V_{\text{low}k}$ interaction shows an even faster convergence than $\VO_{\UCOM}$ and therefore provides a useful reference for a quantitative comparison of different many-body approaches.  

%%%%%%%%%%%%%%%%%%%%%%%%%%%%%%%%%%%%%%%%%%%%%%%%%%%%%%%%%%%%%%%%%%%%%%
\begin{figure}
\includegraphics[width=0.8\columnwidth]{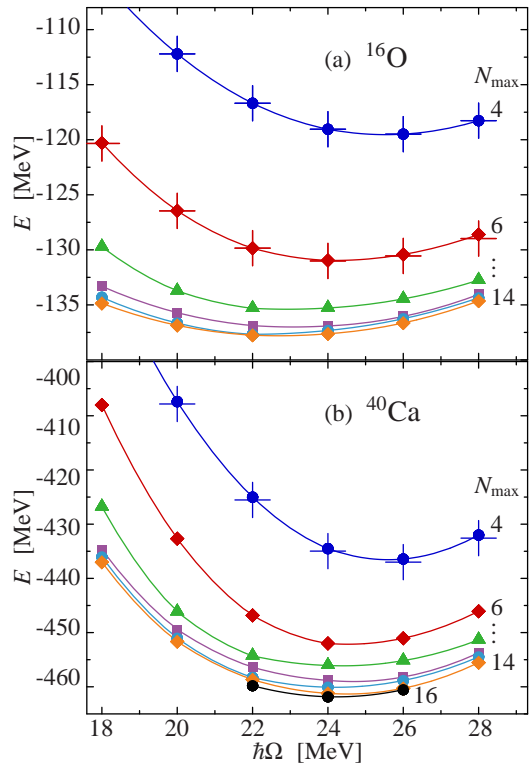}
\caption{(color online) Ground-state energy of \elem{O}{16} (a) and \elem{Ca}{40} (b) as function of the harmonic oscillator frequency $\hbar\Omega$ for different model-space sizes up to $N_{\max}=16$ obtained with the $\VO_{\text{low}k}$ interaction. For \elem{O}{16} up to 4p4h configurations are included, for \elem{Ca}{40} up to 3p3h. Crosses indicate the results of full NCSM calculations.}
\label{fig:O16Ca40_vlowk_Nhwdependence}
\end{figure}
%%%%%%%%%%%%%%%%%%%%%%%%%%%%%%%%%%%%%%%%%%%%%%%%%%%%%%%%%%%%%%%%%%%%%%
%%%%%%%%%%%%%%%%%%%%%%%%%%%%%%%%%%%%%%%%%%%%%%%%%%%%%%%%%%%%%%%%%%%%%%
\begin{table}[b]
\caption{Ground-state energy $E$, point mass radius $R_{\text{rms}}$ and charge radius $R_{\text{ch}}$ obtained for $\elem{O}{16}$ (up to 4p4h for $N_{\max}=14$) and $\elem{Ca}{40}$ (up to 3p3h for $N_{\max}=16$) using the $\VO_{\text{low}k}$ interaction for different oscillator frequencies. $E_{\infty}$ indicates the ground-state energy obtained from an exponential extrapolation.}
\label{tab:O16Ca40_vlowk}
\begin{ruledtabular}
\begin{tabular}{c c c c c c c}
Nucl. & $\hbar\Omega$ [MeV] & $E$ [MeV] & $E_{\infty}$ [MeV] &  $R_{\text{rms}}$ [fm] & $R_{\text{ch}}$ [fm] \\
\hline
\elem{O}{16} & 20 & -136.86 & -137.1 & 2.07 & 2.23 \\
\elem{O}{16} & 22 & -137.75 & -138.0 & 2.03 & 2.20 \\
\elem{O}{16} & 24 & -137.62 & -137.7 & 1.99 & 2.16 \\
\hline
\elem{Ca}{40} & 22 & -459.79 & -461.0 & 2.33 & 2.48 \\
\elem{Ca}{40} & 24 & -461.83 & -462.7 & 2.27 & 2.43 \\
\elem{Ca}{40} & 26 & -460.53 & -461.0 & 2.22 & 2.37\\
\end{tabular}
\end{ruledtabular}
\end{table}
%%%%%%%%%%%%%%%%%%%%%%%%%%%%%%%%%%%%%%%%%%%%%%%%%%%%%%%%%%%%%%%%%%%%%%

The convergence of the ground-state energy of \elem{O}{16} and \elem{Ca}{40} as function of the oscillator frequency $\hbar\Omega$ for different model-space sizes from $N_{\max}=4$ to $14$ is illustrated in Fig. \ref{fig:O16Ca40_vlowk_Nhwdependence}. For \elem{O}{16} configurations up to the 4p4h level have been included, for \elem{Ca}{40} up to 3p3h configurations are considered. For both nuclei we observe a very rapid convergence of the energies, such that fully converged results are obtained. Table \ref{tab:O16Ca40_vlowk} summarizes several ground state observables computed for $N_{\max}=14$ and oscillator frequencies around the energy minimum. In addition to the ground-state energy $E$ for this model space, the energy $E_{\infty}$ resulting from an exponential extrapolation $N_{\max}\to\infty$ is given. For \elem{O}{16} this result can be compared with a recent coupled cluster calculation using the same interaction, which yields an extrapolated energy of $-142.8$ MeV \cite{HagePriv}. This is in good agreement with our result keeping in mind that our calculation provides a variational upper bound, since configurations beyond 4p4h are not included. Furthermore, Tab. \ref{tab:O16Ca40_vlowk} lists the translationally invariant root-mean-square radii (point nucleons) as well as the charge radii (including proton and neutron form factors). From the eigenvectors in the importance truncated space we have also extracted charge form factors and translationally invariant density profiles, which provide an independent means for extracting translationally invariant radii. In comparison with experiment, the two-body interaction $V_{\text{low}k}$ overestimates the binding energies and underestimates the radii significantly. 

In conclusion, we have presented an importance truncation scheme which reduces the model space of the NCSM to the physically relevant states. The relevance of individual basis states is quantified using an \emph{a priori} importance measure derived from many-body perturbation theory. This novel scheme has proven very robust and extends the range of reliable NCSM calculations to regions of the nuclear chart far beyond the p-shell. In this Letter, we have shown the first converged NCSM calculations for the ground state of \elem{Ca}{40} with two different realistic NN interactions. The scheme is universal and can easily be adapted to other quantum many-body problems treated by diagonalization techniques.

%%%%%%%%%%%%%%%%%%%%%%%%%%%%%%%%%%%%%%%%%%%%%%%%%%%%%%%%%%%%%%%%%%%%%%

Supported by the Deutsche Forschungsgemeinschaft through contract SFB 634. This work was partly performed under the auspices of the U. S. Department of Energy by the University of California, Lawrence Livermore National Laboratory under contract No. W-7405-Eng-48. This work was supported in part by the Department of Energy under Grant DE-FC02-07ER41457.

%%%%%%%%%%%%%%%%%%%%%%%%%%%%%%%%%%%%%%%%%%%%%%%%%%%%%%%%%%%%%%%%%%%%%%
%%%%%%%%%%%%%%%%%%%%%%%%%%%%%%%%%%%%%%%%%%%%%%%%%%%%%%%%%%%%%%%%%%%%%%
%%%%%%%%%%%%%%%%%%%%%%%%%%%%%%%%%%%%%%%%%%%%%%%%%%%%%%%%%%%%%%%%%%%%%%

\vfil

%%%%%%%%%%%%%%%%%%%%%%%%%%%%%%%%%%%%%%%%%%%%%%%%%%%%%%%%%%%%%%%%%%%%%%
%%%%%%%%%%%%%%%%%%%%%%%%%%%%%%%%%%%%%%%%%%%%%%%%%%%%%%%%%%%%%%%%%%%%%%
%%%%%%%%%%%%%%%%%%%%%%%%%%%%%%%%%%%%%%%%%%%%%%%%%%%%%%%%%%%%%%%%%%%%%%

\begin{thebibliography}{18}
\expandafter\ifx\csname url\endcsname\relax
  \def\url#1{\texttt{#1}}\fi
\providecommand{\eprint}[2][]{\url{#2}}

\bibitem{WiSt95}
 R.~B. Wiringa, V.~G.~J. Stoks, and R.~Schiavilla,
 Phys. Rev. C \textbf{51}, 38 (1995).

\bibitem{Mach01}
 R.~Machleidt,
 Phys. Rev. C \textbf{63}, 024001 (2001).

\bibitem{EnMa03}
 D.~R. Entem and R.~Machleidt,
 Phys. Rev. C \textbf{68}, 041001(R) (2003).

\bibitem{EpNo02}
 E.~Epelbaum \textit{et al.}, 
 Phys. Rev. C \textbf{66}, 064001 (2002).

\bibitem{NaVa00}
 P.~Navr\'atil, J.~P. Vary, and B.~R. Barrett,
 Phys. Rev. Lett. \textbf{84}, 5728 (2000).

\bibitem{NaOr02}
 P.~Navr\'atil and W.~E. Ormand,
 Phys. Rev. Lett. \textbf{88}, 152502 (2002).

\bibitem{PiWi01}
 S.~C. Pieper and R.~B. Wiringa,
 Ann. Rev. Nucl. Part. Sci. \textbf{51}, 53 (2001).

\bibitem{PiWi04}
 S.~C. Pieper, R.~B. Wiringa, and J. Carlson,
 Phys. Rev. C \textbf{70}, 054325 (2004).

\bibitem{NaGu07}
 P.~Navr\'atil \textit{et al.}, (2007), \eprint{nucl-th/0701038}.

\bibitem{RoPa06}
 R.~Roth \textit{et al.}, Phys. Rev. C \textbf{73}, 044312 (2006).

\bibitem{BaPa06}
 C.~Barbieri \textit{et al.}, (2006), \eprint{nucl-th/0608011}.

\bibitem{KoDe04}
 K.~Kowalski \textit{et al.}, Phys. Rev. Lett. \textbf{92}, 132501 (2004).

\bibitem{WlDe05}
 M.~W\l{}och \textit{et al.}, 
 Phys. Rev. Lett. \textbf{94}, 212501 (2005).

\bibitem{CaMa05}
  E.~Caurier \textit{et al.}, Rev. Mod. Phys. \textbf{55}, 427 (2005).

\bibitem{ScHi07}
  F.~Schmitt, M.~Hild, and R.~Roth, 
  J. Phys. B: At. Mol. Opt. Phys. \textbf{40}, 371 (2007).

\bibitem{RoHe05}
  R.~Roth \textit{et al.}, 
  Phys. Rev. C \textbf{72}, 034002 (2005).

\bibitem{BoKu03} 
  S.K. Bogner, T. T. S. Kuo, and A. Schwenk, 
  Phys. Rept. \textbf{386}, 1 (2003).

\bibitem{CaNo99}
  E.~Caurier and F.~Nowacki,
  Acta Phys. Polonica \textbf{30}, 705 (1999).

\bibitem{HagePriv}
  G. Hagen, private communication.

\end{thebibliography}
\end{document}